\documentclass{mn2e}

\renewcommand\[{\begin{equation}}
\renewcommand\]{\end{equation}}

\def\kpc{\,{\rm kpc}}

\def\K{\,{\rm K}}
\def\msun{{\rm M_{\odot}}}
\def\kms{\,{\rm km}\, {\rm s^{-1}}}

\def\d{{\rm d}}

\def\i{\relax\ifmmode{\rm i}\else\char16\fi}

\def\lesssim{{_ <\atop{^\sim}}}
\def\lta{\lesssim}
\def\grtsim{{_ >\atop{^\sim}}}
\def\gta{\grtsim}

\def\lesssim{\mathrel{\hbox{\rlap{\hbox{\lower4pt\hbox{$\sim$}}}\hbox{$<$}}}}
\def\gtrsim{\mathrel{\hbox{\rlap{\hbox{\lower4pt\hbox{$\sim$}}}\hbox{$>$}}}}

\def\apj#1 #2{ApJ, #1, #2}
\def\aj#1 #2{AJ, #1, #2}
\def\mn#1 #2{MNRAS, #1, #2}
\def\aa#1 #2{A\&A, #1, #2}

\begin{document}

   \title[The galaxy luminosity function]
   {On the origin of the galaxy luminosity function}

   \author[J. Binney]
          {James Binney
           \\
           Theoretical Physics, Keble Road, Oxford OX1 3NP\\
          }

   \date{}

   \maketitle

\begin{abstract}
Evidence is summarized which suggests that when a protogalaxy collapses, a
fraction $f$ of its gas fails to heat to the virial temperature, where $f$ is
large for haloes less massive than the value $M^*$ associated with $L^*$
galaxies. Stars and galaxies form only from the cool gas fraction. Hot gas is ejected
from low-mass systems as in conventional semi-analytic models of galaxy
formation. In high-mass systems it is retained but does not cool and form
stars. Instead it builds up as a largely inert atmosphere, in which cooling
is inhibited by an episodically active galactic nucleus. Cold gas
frequently falls into galactic haloes. In the absence of a dense atmosphere
of virial-temperature gas it builds up on nearly circular orbits and can be
observed in the 21 cm line of HI. When there is a sufficiently dense hot
atmosphere, cold infalling gas tends to be ablated and
absorbed by the hot atmosphere before it can form stars.
The picture nicely explains away the surfeit of high-luminosity galaxies
that has recently plagued semi-analytic models galaxy formation, replacing
them by systems of moderate luminosity from old stars and large X-ray
luminosities from hot gas.
\end{abstract}


\section{Introduction}

The Cold Dark Matter theory of structure formation has enjoyed notable
successes in recent years. Not only does it provide a unified interpretation
of the cosmic microwave background (CMB) and galaxy clustering, but on
smaller scales it has led to successful models of the Ly$\alpha$ forest and
gravitational lensing by galaxies, both strong and weak. While
controversy surrounds the compatibility of the CDM theory with the measured
dark-matter densities deep within galaxies, it seems likely that problems in
this area arise from our poor understanding of galaxy formation rather than
pointing to genuine shortcomings in the CDM model (Binney 2003).

Since the CDM theory is a theory of the invisible, tests of the theory rely
to a large extent on predictions of how dark matter affects the baryon
content of the Universe. On large scales forces other than gravity can
generally be neglected, so the physics is simple and reliable. On the
smallest scales both strong and electromagnetic interactions between baryons
are as important as gravitational interactions, so the physics is complex.
We know that this physics somehow gives rise to stars and galaxies, but the
details are obscure.

Given the importance of galaxy-formation theory for making observational
predictions from the CDM model, strenuous efforts have been made to model
galaxy formation in the presence of CDM. These efforts are strongly
influenced the papers of Rees \& Ostriker (1977) and White \& Rees (1978),
which took it as axiomatic that gas that falls into a gravitational
potential well immediately shocks to the well's virial temperature. Galaxies
were assumed to form as this rapidly heated gas slowly cooled.  I argue here
that this picture is wrong. In reality galaxies are formed from the the
significant fraction of gas that fails to heat as it falls into the
potential well, and that the fraction that immediately heats to the virial
temperature is unable to cool subsequently. As the clustering hierarchy of
the CDM model develops, the fraction of gas that heats to the virial
temperature increases, so the efficiency of galaxy formation falls.  The
argument draws on recent developments in the study of `cooling flows',
simulations of baryonic infall (Birnboim \& Dekel 2003; Katz et al.\ 2003),
and semi-analytic galaxy-formation theory (Benson et al.\ 2003).

\section{Does gas virialize?}

The influential papers of galaxy-formation theory by Rees \& Ostriker (1977)
and White \& Rees (1978) considered it evident that when gas falls into a
potential well, it would be shock heated to the well's virial temperature.
In my thesis work (Binney 1977) I studied the shock heating of infalling gas
in circumstances that maximized the chances of the gas shock heating:
inspired by Zel'dovich's pancake theory of galaxy formation I considered the
accumulation of gas in a sheet of uniform surface density following the
collapse of an initially homogeneous spheroidal body of gas. Gas in the
equatorial plane was stationary at all times but increased in density as gas
fell onto it from the polar directions with steadily increasing speed. As
the speed of the infalling gas increased, the post-shock temperature of the
infallen gas rose. In analytic work I compared the cooling time of the
post-shock gas with the time required for infalling gas to travel a distance
equal to the distance of the shock from the equatorial plane. This
comparison showed that if the cooling rate was dominated by bremsstrahlung,
the shock would break away from the equatorial plane only if the radius of
the collapsing ellipsoid exceeded $30\kpc$. I used one-dimensional
hydrodynamical simulations to test the analytical work and found that it
seriously understimated the importance of cooling because (i) the infall
velocity only gradually rises to the characteristic virial value, and (ii)
even in a pure hydrogen-helium plasma, cooling at low temperatures greatly
exceeds the bremsstrahlung rate. In a galaxy-sized collapse bremsstrahlung
never dominates the cooling rate, and even for a cluster-sized collapse
bremsstrahlung is important only when the last mass fraction falls in.
Consequently, the fraction of gas that heats to the virial temperature is
negligible in a galaxy-sized collapse and much less than unity in a
cluster-sized one.

I considered the impact on the heated gas fraction of inhomogeneities in the
infalling gas, and concluded that they were likely to favour cooling by
increasing the surface density at which gas impacts the equatorial plane.
Hence I concluded that essentially no gas would heat to the virial
temperature when a galaxy-sized halo collapses, in flat contradiction to the
fundamental premise of the influential paper of Rees \& Ostriker (1977).

When CDM replaced neutrinos as the favoured dark matter component in the
1980s, it became necessary to revisit my 1977 study because CDM gives rise
to smaller and cuspier potential wells, favouring cooling, while it enables
one to lower the baryon density, which favours heating. Demoralized by the
scant attention paid to my 1977 study I did not do so, but recently Birnboim
\& Dekel (2003) have done something similar. They have performed
one-dimensional hydrodynamical simulations of baryonic infall into a
spherical potential well, and concluded that ``most galactic haloes that
have collapsed and virialized by $z\sim2$ did not produce a virial shock.
Haloes less massive that $10^{11}\msun$ never produce a shock even if the
gas has zero metallicity. If the metallicity is non-negligible (e.g.,
$Z\sim0.05$) this lower bound to shock formation rises to
$\sim7\times10^{11}\msun$.'' It seems that the bottom line has changed very
little in 25 years.

Further support for my contention that infall is ineffective in heating gas
to the virial temperature is provided by Katz et al.\ (2003). They used
smooth-particle hydrodynamics (SPH) to follow the evolution of a mixture of
CDM and baryons in cosmological clustering simulations. They measured the
maximum temperature $T_{\rm max}$ reached by each gas particle within a
given halo, and found that at infall redshifts $z\gta1$ the distribution of
$T_{\rm max}$ is distinctly bimodal, with about equal masses associated with
peaks centred on $\sim 3\times10^4\K$ and the virial temperature. At smaller
values of $z$, the gap in a histogram of $T_{\rm max}$ values fills in, but
the distribution of $T_{\rm max}$ values remains extremely broad, so a great
deal of matter continues to fall in without heating to near the virial
temperature. The quantity of infalling gas diminishes strongly over time, so
most gas falls in when the temperature distribution is strongly bimodal.

Future increases in the spatial resolution of the simulations is likely to
increase the fraction of gas that falls in  cold since increasing the
resolution raises the density of the densest, fastest-cooling gas and thus
makes it harder for gas to heat to the virial temperature.

\section{Feedback and the galaxy luminosity function}

The clustering of CDM has been exstensively simulated and the popular
`semi-analytic model' of galaxy formation has been used to associate with
each dark-matter halo a galaxy of given luminosity and colours (Kauffmann,
White \& Guiderdoni, 1993; Lacey et al.\ 1993; Cole et al.\ 1994; Kauffmann
et al.\ 1999; Somerville \& Primack 1999; Cole et al.\ 2000; Benson et al.\
2003). The algorithms employed in semi-analytic models
start by assuming that each halo's quota of baryons shock heats to the
virial temperature and then cools on the timescale that follows from the
density of the gas. As gas cools it forms a cold disk within which stars
form at some rate, so the mass and age
distribution of the stellar population of any halo is determined, and its
luminosity and colours can be predicted. Cole et al.\ (2000) provide a lucid
account of this  very detailed work.

The mass function of dark-matter haloes $\Psi(M)=\d N/\d M$ is predicted by
simulations of dark-matter clustering (Jenkins et al.\ 2001). It is quite
different from the galaxy luminosity function $\Phi(L)=\d N/\d L$ in two
respects: (i) while at the low-$M$ end $\Psi(M)$ is rather close to a power
law $\Psi(M)\propto M^{-\beta}$ with $\beta\sim 2.17$, the luminosity
function of galaxies can be fitted by Schechter's (1976) formula
$\Phi(L)\propto L^{-\alpha}\exp(-L/L^*)$ with a flatter slope:
$\alpha\sim0.95$ (Cole et al.\ 2001; Kochanek et al.\ 2001). (ii) While both
$\Psi$ and $\Phi$ fall below these power laws at large values of their
argument, the luminosity $L^*$ of the break in $\Phi(L)$ corresponds to
galactic masses that are $\gta100$ times smaller than the mass of the break
in $\Psi$.  Thus, if one predicted $\Phi(L)$ by converting $\Psi(M)$ using
the mass-to-light ratio of a typical $L^*$ galaxy, one would predict too
many galaxies at both extremes of the luminosity scale.

Two mechanisms provide plausible explanations of the relative
dearth of low-luminosity galaxies: upon reionization at $z_{\rm ion}\sim6$,
the temperature of the baryons would have risen to $T\sim2\times10^4\K$ even
before any shock heating. Gas at this temperature cannot be effectively
trapped in a potential well with a peak circular velocity $v_{\rm max}$
smaller than $\sim35\kms$ (Efstathiou 1992; Quinn Katz \& Efstathiou 1996;
Thoul \& Weinberg 1996).  Since only the more massive haloes have larger
circular speeds, the only low-mass haloes with stars will be those that
formed before $z_{\rm ion}$. This prediction chimes nicely
with the observation that dwarf galaxies are overabundant in clusters of
galaxies because on the average cluster galaxies formed before comparable
field galaxies.

Another mechanism capable of explaining the dearth of low luminosity
galaxies is feedback from supernovae: a burst of star formation is expected
to initiate the formation of a galaxy. The massive stars in this burst will
complete their lives and explode as core-collapse supernovae on a timescale
that is short compared to the dynamical time of the embedding halo. The
supernova blasts will heat surrounding gas to temperatures $\gta10^6\K$. If
$v_{\rm max}$ is smaller than $\sim100\kms$, the hot gas will stream outwards
and further star formation may be effectively suppressed (Dekel \& Silk 
1986).

Benson et al (2003) have recently re-examined  the fit to the galaxy
luminosity function that one obtains by applying the semi-analytic model to
the known mass function of dark-matter haloes. They conclude that with the
relatively large baryon density that is now mandated by studies of
large-scale structure, especially the CMB, the semi-analytic model cannot
simultaneously fit both the low- and the high-luminosity end of $\Phi(L)$.
The strength of feedback by supernovae is poorly constrained and may be
treated as a free parameter. When strong feedback is chosen, the density of
low-luminosity galaxies can be made sufficiently small. However, the baryons
expelled from low-luminosity systems later fall into haloes of larger mass
and make them more luminous. The consequence is that there are then too many
high-luminosity galaxies.

\section{Cooling flows}

We observe systems that contain gas at the virial temperature: X-ray
observations long ago showed that nearly all clusters of galaxies and most
luminous elliptical galaxies have such gas. Thus we are in a position to
check observationally an important ingredient of the semi-analytic model
of galaxy formation.

The cooling time of the gas in galaxy clusters and elliptical galaxies
decreases as one approaches the centre, and in most systems the central
value of the cooling time is much shorter than the Hubble time. This
observation led to the expectation that gas in these systems was steadily
forming stars as it cooled (Cowie \& Binney 1977; Fabian \& Nulsen 1977).
From the radial surface-brightness profile of the X-rays and the assumption
that the `cooling flow' had reached a steady state, one can derive the rate
of star formation as a function of radius (Nulsen 1986).  Massive-star
formation is certainly not taking place at the rate predicted by a
Salpeter-like initial mass function (IMF), and formation of predominantly
low-mass stars could also be excluded (Prestwich et al.\ 1997). Nor was cold
gas present in the quantity required if star formation was somehow
suppressed (Donahue et al.\ 2000; Edge 2001; Edge et al.\ 2002). Data from
the Chandra and XMM-Newton satellites has recently shown that very little if
any gas is cooling in clusters of galaxies, notwithstanding their short
cooling times (Tamura et al.\ 2001; Peterson et al.\ 2002). Appropriate data
are not yet available for individual elliptical galaxies, but the
presumption must be that their smaller `cooling flows' behave in a similar
way. It is evident that the gas is being heated. 

The details of how the gas is heated are
controversial and irrelevant to the present discussion. What matters is that
(a) very little if any gas is cooling to temperatures lower than $\sim1/3$
that of the main body of gas, (b) any cold gas is confined to the very
centre of the system (Edge et al. 2002), and (c) this gas is dusty and is quite
likely material that has fallen in cold and is unconnected with the hot gas
(Sparks et al.\ 1989). 

Another important implication of the study of extended X-ray sources is that
the gas observed in these systems is not heated to the virial temperature by
gravitational shocking, but by supernovae and/or an active galactic nucleus
(AGN). This fact was first recognized by Kaiser (1991), who pointed out that
in the contrary case not only would the number of bright X-ray sources at
$z\sim1$ be greater than observed, but the correlation between luminosity
and temperature at the present epoch would be less steep than is observed
(Kaiser 1986).
Kaiser's inference has been confirmed in many subsequent studies (e.g.,
Shimizu et al.\ 2003).
Direct evidence for non-gravitational heating is provided by the large
fraction ($\gta50$ percent) of the heavy elements in clusters of galaxies
which are contained in the X-ray emitting gas but must have been synthesized
within the galaxies.

\section{Calling a halt to star formation}

The study of `cooling flows' teaches us that systems in which all gas is at
the virial temperature do {\it not\/} form stars and do {\it not\/}
accumulate cold gas from which stars could form at a later date. The
evidence for cold infall suggests that star formation in such a system could
be restarted if a quantity of cold gas fell in. An episode of cold infall
will not necessarily lead to star formation, however, because a sufficiently
dense hot atmosphere will ablate and reheat infalling gas before it can form
stars. In fact, we can assume that cold gas enters the virialized system at
a density that does not permit rapid star formation, since otherwise it
would already have fragmented into stars. Hence to form stars it has to
accumulate in the halo potential, probably on some approximately closed
orbits. The extended HI disks that are frequently observed around late-type
galaxies are regions where such accumulation of cold infalling material is
taking place. A sufficiently dense, hot atmosphere will inhibit this
accumulation by ablating and stripping the cold gas, both (i) as it falls
towards the closed parking orbits, and (ii) within those orbits. It will be
highly vulnerable to ablation in phase (i) because its surface density will
be low, but stage (ii) can last longer. A credible ab initio calculation of
the ablation rate is hard because it would require a quantitative
understanding of the magnetized, turbulent boundary layer between the cold
and hot gas, which differ by at least three orders of magnitude in density.

\section{Putting it all together}

The following scheme seems to account neatly for all the evidence summarized
above. When a dark halo virializes, a fraction $f$ of the progenitor's gas
remains at $T\lta10^4\K$, while the remainder heats to the virial
temperature. The fraction $f$ decreases as the halo mass increases. If the
halo is massive enough, the fraction $f$ is retained as a disk in which star
formation commences. The gravitationally heated fraction is further heated
by dying stars and by any massive black hole that forms at the system's
nucleus. In a lower-mass halo this gas expands and flows out of the system.
As we move up the sequence to larger masses, the temperature that must be
reached to push the hot gas out increases, and at some mass $M^*$, which we
must identify with that of $L^*$ galaxies, the gas cannot be pushed
completely out of the halo, although it can be pushed so high into the halo
that it takes several gigayears to fall back once the rate of heating by
core-collapse supernovae has moderated (D'Ercole et al 1989). Once the hot
gas cannot escape, its density steadily increases as dying stars inject
fresh hot gas. At some point it becomes dense enough to ablate and absorb
any infalling cold material before it can give rise to significant star
formation.

The density of virial-temperature gas increases rapidly with halo mass near
$M^*$ because both the fraction $1-f$ and the ability to retain
supernova-heated gas increase with $M$.

Current semi-analytic models of galaxy formation form too many
high-luminosity galaxies when appropriate feedback is used because they
assume that gas heated to the virial temperature can cool and make stars.
Once this error is corrected, the models will yield luminosity functions
that cut off sharply above $L^*$. Instead the revised models will predict
that high-mass haloes will contain large masses of X-ray emitting virialized
gas, in agreement with observation.

\section{Conclusions}

Observations of cooling flows lead to the conclusion that virial-temperature
gas cannot cool and give rise to star formation. The inability of hot gas to
cool when supernova heating is effective is obvious. The continued inability
of this gas to cool once the supernova heating fades has surprised many, but
can be readily explained and was predicted (Tabor \& Binney 1993; Binney \&
Tabor 1995). The key point is that any very cold gas that forms, does so in
the immediate viscinity of the central black hole, and induces accretion
that promptly reheats the gas, largely shutting off further accretion. The
details of this chain of events are inadequately understood, but the
prospects for elucidating them are good (Reynolds et al, 2001; Omma et al 03).

Since gas heated to the virial temperature cannot cool and form stars,
galaxy formation would be impossible if Rees \& Ostriker were correct in
assuming that gas is heats to the virial temperature when a system
virializes. Three different simulations argue to the contrary that a
significant fraction of the gas never reaches the virial temperature.
Galaxies form from this under-heated fraction. In lower-mass systems the
heated fraction is largely ejected. Some of this subsequently finds its way
into more massive haloes. These haloes heat a larger fraction of their
progenitor's gas to the virial temperature and tend to retain both this gas
and gas ejected by their stellar systems. Consequently, they develop dense
atmospheres of virial-temperature gas. These atmospheres ablate any cold gas
that falls into their haloes, so star formation ceases in these systems.
Consequently, high-mass haloes are now associated with only modest
luminosities from rather old stars and substantial masses of X-ray emitting
virial-temperature gas.

\end{document}